\documentclass[useAMS,usenatbib]{mn2e}

\usepackage{epic}
\usepackage{epsfig}
\usepackage{rotating}
\usepackage{url}

\title[Correlation of supernova redshifts with CMB]{Correlation of 
supernova redshifts with temperature fluctuations of the Cosmic Microwave Background}
\author[V. N. Yershov, V. V. Orlov  and A. A. Raikov]{V. N. Yershov$^{1}$\thanks{E-mail:
vny@mssl.ucl.ac.uk},
V. V. Orlov$^{2,3}$ and A. A. Raikov$^3$\\
$^1$Mullard Space Science Laboratory, University College London, 
Holbury St.Mary, Dorking, RH5 6NT U.K.\\
$^2$Saint Petersburg State University, 
28 Universitetskij prospect, Peterhof,  
198504, St.Petersburg, Russia\\
$^3$Main (Pulkovo) Astronomical Observatory 
of the Russian Academy of Sciences, 65 Pulkovskoe shosse, 
196140, St.Petersburg, Russia}

\begin{document}

 \date{Accepted ---. Received ---; in original form ---}

\maketitle

\begin{abstract}
Redshifts of a supernova (SN) and gamma-ray burst (GRB) samples are compared
with the pixel temperatures of the {\it Wilkinson Microwave Anisotropy Probe (WMAP)} 
seven-years data, the pixels locations corresponding
to the SN and GRB sky coordinates. We have found 
a statistically significant correlation of the SN 
redshifts with the WMAP data, the average temperature deviation 
being $+29.9\pm4.4~\mu{\rm K}$ for the redshifts $z$ ranging from
0.5 to 1.0  and $+8.6\pm 1.3~\mu{\rm K}$ for $z\in(0.0,0.4)$.
The latter value accords with the theoretical estimates for the 
distortion of the cosmic microwave background
due to the integrated Sachs-Wolfe effect, whereas the larger anomaly
for higher redshifts should be studied in more detail in the future.   
\end{abstract}


DOI:10.1111/j.1365-2966.2012.21026.x

\section{Introduction}
During the past few years there has been a noticeable increase in 
publications discussing different aspects of
non-Gaussian features in the cosmic microwave 
background (CMB) radiation 
\citep{ferreira98,wu01,verde01,komatsu03,park04,vielva04,%
troya07,vachurin09,bernui10,pogosyan11,ashoorioona11}. 
The non-Gaussian behaviour seen on different angular scales and with a variety 
of amplitudes imply the existence of some large-scale anisotropies 
in the distribution of matter in the Universe. 

One of such anisotropies is a so far unexplained alignment 
of low-order multipoles  ($l=2$ to $5$) in the representation of CMB
as a power series expansion with spherical harmonics.
This anomaly is sometimes referred to as ``the cosmic axis of evil'' 
(\citealt{spergel03,deoliveira04,schwarz04,land05,magueijo07}; 
see also the review 
by \citealt{rakic07}).
Furthermore, recently it was found that the axes of rotation 
of most galaxies from the Sloan Digital Sky Survey (SDSS) appear to line 
up with the axis of evil \citep{longo11}. 
The other known anomalies are related to the so-called ``hot'' and ``cold''
spots of the CMB  \citep{larson04,larson05,vielva07,copi10,ayaita10} 
which are regions of increased or reduced temperatures
compared to the average temperature of the CMB map. 
For example, the temperature of one of the most prominent 
and broadly discussed non-Gaussian features known as the 
extreme cold spot 
\citep{vielva04,mukhrjee04,martinez06,cruz07, vielva10} 
is approximately 70~$\mu$K lower 
than the average CMB temperature.  
The possibilities of such a large anomaly appearing 
within the framework of the standard $\Lambda$-cold-dark-matter
($\Lambda$CDM) model
are discussed by \citet{inoue12}, who argues that there might 
exist a large-scale void along the 
line of view towards the extreme cold spot.  
The existence of a void of 100~Mpc radius along this line 
at $0.5<z<0.9$ was recently ruled out by \citet{granett10}, who used
galaxy counts in photometric redshift bins. 
However, the void might be at larger distances.
The non-uniform distribution of the largest CMB spots
over the celestial sphere results in the north-south ecliptic 
asymmetry in the CMB maps (\citealt{eriksen04,bernui08}; 
 see also the discussion in \citealt{rakic07}).  

A variety of techniques are used for extracting the information
about CMB anisotropies, such as the determination of
both angular and planar modulation parameters
by using the two-point correlation function \citep{szapudi01}, 
the characterisation of deviations 
from a given (say, Gaussian) model using the $\chi^2$-criterion,
or the comparison of random Gaussian maps with 
observed maps \citep{liguori03, liguori07}.
 
Also of importance is the interpretation of the deviations 
from a Gaussian distribution found in the CMB power spectrum.
One of the possible explanations of these deviations 
is based on the assumption that they might be due to 
contamination of CMB by foreground sources. The induced deviations 
caused by the strongest of
these sources are removed by the foreground reduction procedure, although 
the completeness of this procedure is still under scrutiny 
\citep[see][]{naselsky03,then06,bennett11}.

It is worthwhile mentioning the 
 Sachs-Wolfe \citep{sachs67}, 
 Rees-Sciama \citep{rees68} and Sunyaev-Zeldovich \citep{sunyaev70} effects. 
The first two alter the energy of CMB photons
when they traverse gravitational fields of different 
strengths corresponding to either voids or conglomerations of matter. 
In $\Lambda$CDM models, the ``late-time'' Sachs-Wolfe effect 
works mostly on large angular scales \citep{bartolo10}. 
The third is related to the scattering of
CMB photons from high energy electrons in hot coronae of
galaxy clusters. 

The distinction between the proper and induced CMB anomalies 
can be made by comparing the WMAP data with patterns 
created by other sources on the sky.
For example, \citet{verkhodanov10} have found 
statistically significant correlations between
CMB map temperatures and the distribution of
gamma-ray bursts (GRB). 
Significant correlations of WMAP data 
with galaxy samples on large angular scales of $4^\circ$ to 
$10^\circ$ were found by \citet{fosalba04} for 
the Automated Plate Measuring galaxy survey, 
by \citet{fosalba03,cabre06} for the Sloan Digital Sky Survey,
by \citet{goto12} for the wide-field infrared survey 
of galaxies and by other authors. 

Since there exists a broad variety of opinions 
with respect to the nature of the anomalies seen in the WMAP data,
further statistical studies
of these data by using alternative techniques are desirable.
In particular, it is important to continue with
cross-correlations of CMB maps with spatial distributions of different  
high-redshift objects, such as galaxies, quasars
or GRBs.

Most of the cross-correlations are based on two-dimensional distributions
of extragalactic sources. Here we shall explore the third coordinate 
(redshift $z$) for checking whether some of the CMB fluctuations might be caused by 
distant concentrations of matter. 
We shall use supernovae (SNe) and GRB samples
as the test objects that trace matter in the universe.


\section{CMB temperature as a function of SN redshifts}

For our analysis we have taken the list of supernovae \citep{bartunov07} 
compiled by the Sternberng Astronomical Institute (Moscow) 
at \url{http://www.sai.msu.su/sn/sncat} 
which contains a large sample of SNe with known redshifts
(5688 SNe for September 2011).
We have also used sources from 
the GRB list maintained by J. Greiner \citet{greiner08} at Max-Planck-Institut
  f\"ur extraterrestrische Physik (Garching) at 
\url{http://www.mpe.mpg.de/~jcg/grbgen.html}. With the updates from 
 \citet{cucciara11} and the Swift data archive at 
\url{http://www.swift.ac.uk} this list 
contained 992 GRBs (as for October 2011).

The WMAP data products \citep{wmap10} were obtained from the NASA archive
at \url{http://lambda.gsfc.nasa.gov}, and the software for 
processing these data was adopted from the comprehensive
HEALPix package developed by Jet Propulsion Laboratory and
available at \url{http://healpix.jpl.nasa.gov} \citep{gorski05}.
Since most of the correlations between the CMB and extragalactic
sources are found on large angular scales, we have used here
the foreground reduced WMAP sky maps with the resolution 
corresponding to the HEALPix parameter $N_{\rm side}=128$.  
\begin{figure}
\hspace{-0.5cm}
\epsfig{figure=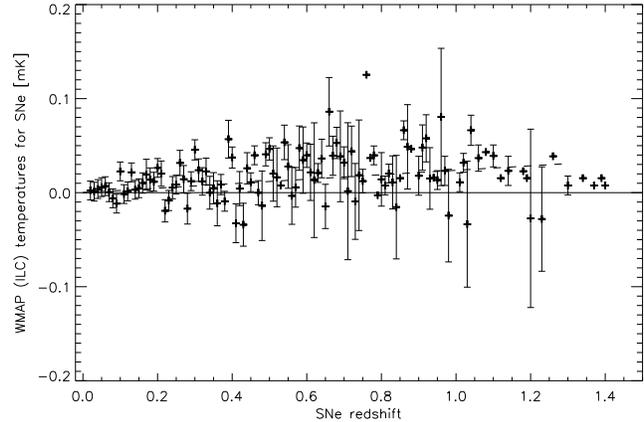,width=9.0cm}
\vspace{-0.2cm}
\caption{Histogram of the WMAP pixel temperatures $T_{\rm SN}$ near
the supernovae locations   
as a function of SN redshifts $z_{\rm SN}$ (for the WMAP ILS 
values $T_{\rm SN}$); 
the points with no error-bars correspond to 
the redshift bins containing a single
source. The dashed line  indicates the slope of the 
linear regression of the SN temperatures: 
$T_{\rm SN}=(+23.7\pm6.0) \,z_{\rm SN} ~\mu{\rm K}$}. 
\label{fig:sn_temp_redshift}
\end{figure}

Figure~\ref{fig:sn_temp_redshift} shows the histogram of the 
WMAP pixel temperatures for the SNe locations as a function 
of SNe redshifts for the WMAP Internal Linear Combination (ILC) map
with the redshift ranging from zero to 1.4.

\begin{figure}
\vspace{-0.7cm}
\begin{turn}{90}
\epsfig{figure=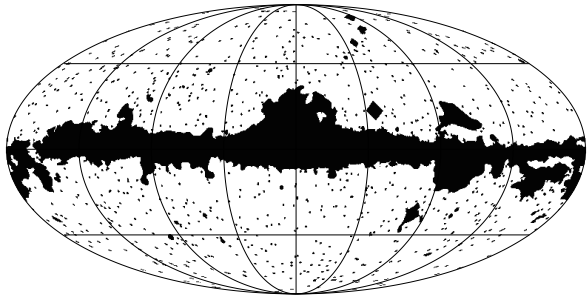,width=4.9cm}
\end{turn}
\vspace{-0.45cm}
\caption{WMAP temperature analysis mask 
applied to the WMAP ILC-map prior the computation of
the SN and GRB redshift histograms.} 
\label{fig:temp_mask}
\end{figure}

In order to avoid contamination from point sources and 
from the Galactic plane, we have used  
the WMAP temperature analysis mask \citep{wmap10}
which includes point sources from 
external catalogues, as well all as the proper WMAP point source
catalogue. It also takes into account the galactic foreground 
emission \citep{gold11}. 
We have taken the temperature analysis mask with higher resolution 
corresponding to $N_{\rm side}=512$ (Figure~\ref{fig:temp_mask})
and applied it to the same higher resolution ILC-map, 
after which we degraded the masked ILC-map
to the lower resolution of $N_{\rm side}=128$.

\begin{figure}
\vspace{-0.7cm}
\begin{turn}{90}
\epsfig{figure=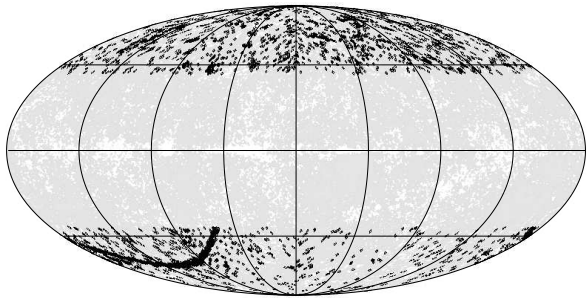,width=4.9cm}
\end{turn}
\vspace{-0.45cm}
\caption{Sky distribution of SNe for the selected range of 
latitudes $|b|>40^\circ$. The strip showing high SN densities 
corresponds to the SDSS Supernova Survey covering the  
equatorial strip of approximately $2.5^\circ$-width}. 
\label{fig:sn_distr1}
\end{figure}

Additionally, we have restricted our SN and GRB samples 
to the galactic latitudes
$|b\,|>40^\circ$ and rejected the 
WMAP pixels with temperature values $|T| >0.5$~mK, as such pixels 
are likely to correspond to artifacts or some extra point sources.
The distribution of SNe over the sky within the selected
latitude limits is shown in Figure~\ref{fig:sn_distr1}.

As for the GRB data, only part of them contains photometric 
redshifts, so we included in our analysis the GRB 
pseudo-redshifts from the catalogue of  \citet{pelangeon08} available 
at \url{http://cosmos.ast.obs-mip.fr}
(the empirical methods for estimating pseudo-redshifts are described 
by \citealt{atteia03,band04}).

By grouping the WMAP pixel temperatures into redshift bins
with the bin size of 0.2 for GRBs and 0.01 for SNe,
we have built two histograms for examining the possible relationship between 
the SN and GRB redshifts and CMB temperatures.

\begin{figure}
\hspace{-0.5cm}
\epsfig{figure=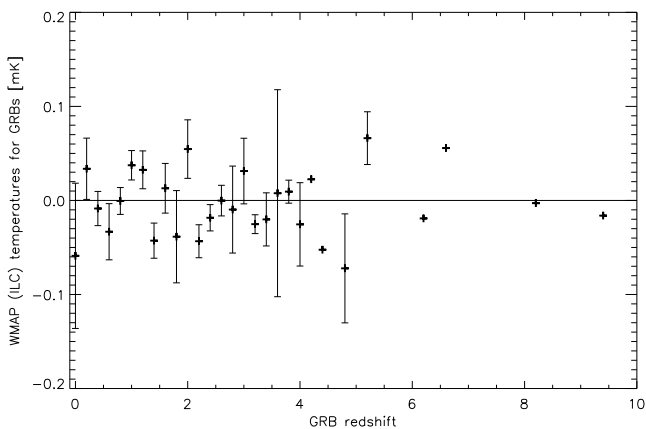,width=9.0cm}
\vspace{-0.7cm}
\caption{Histogram of the WMAP pixel temperatures $T_{\rm GRB}$ near the 
GRB locations on the sky as a function of redshifts $z_{\rm GRB}$ (for the 
WMAP ILC values $T_{\rm GRB}$); the points with no error-bars correspond 
to the redshift bins containing a single source.} 
\label{fig:grb_temp_redshift}
\end{figure}

The histogram in Figure~\ref{fig:sn_temp_redshift} shows 
the bin-averaged values of the WMAP temperatures $T_{\rm SN}$
for supernovae; the values $T_{\rm GRB}$ for gamma-ray bursts
are shown in Figure~\ref{fig:grb_temp_redshift}. 
The error-bars in these plots indicate the standard errors
(SE) of the averages $T_{\rm SN}$ (or $T_{\rm GRB}$)
in each redshift bin calculated as
$SE=\sqrt{\sum_{i=1}^n (T_{\rm SN}^i-T_{\rm SN})^2/n(n-1)}$,
where $n$ is the number of SNe in the bin. The points corresponding to a single source
in the bin are shown having no error-bars.  
The average WMAP pixel temperature
for the selected latitude range, 
$\overline{T}_{|b|>40^\circ}=+2.7 \pm 0.2 ~\mu{\rm K}$, was 
subtracted from the bin-averages 
(the positive shift of $\overline{T}_{|b|>40^\circ}$ is 
due to the latitude selection effect, as the zero-average corresponds
to the whole sky only).
Although the data are noisy,
one can notice a statistically significant
 deviation of $T_{SN}$ from the average temperature 
$\overline{T}_{|b|>40^\circ}$, which is larger for  
the redshifts ranging from 0.5 to 1.0, where 
the average of $T_{\rm SN}$ is
$\overline{T}_{\rm SN}^{(0.5,1.0)}=+29.9\pm4.4~\mu{\rm K}$,
and which is smaller for lower redshifts $z\in(0.0,0.4)$:
$\overline{T}_{\rm SN}^{(0.0,0.4)}=+8.6\pm 1.3~\mu{\rm K}$, 
the average number of SNe in each redshift bin 
being 52 for $z\in(0.0,0.4)$ and 6.5 for $z\in(0.5,1.0)$. 
For the whole sample of 
2413 SNe within the selected latitude range, the average temperature is
also significantly positively biased: $\overline{T}_{\rm SN}=+12.0\pm1.6~\mu{\rm K}$.
Such an excess of $\overline{T}_{\rm SN}$ 
is larger than the current theoretical and 
observational estimates for the potential
biases in the CMB temperatures due to extragalactic
contamination \citep{knox98,santos03,zahn05,serra08,taburet09}.

\begin{figure}
\hspace{-0.5cm}
\epsfig{figure=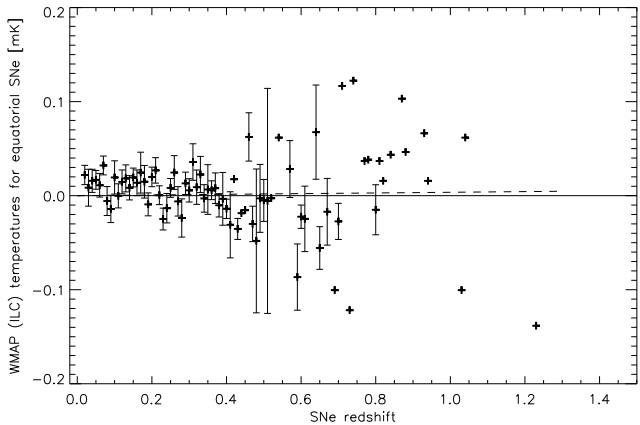,width=9.0cm}
\caption{Histogram $T_{\rm SN}/z_{\rm SN}$ for the supernova subset within the
equatorial latitudes $\pm 1.5^\circ$ seen as a high-SN-density strip in
Figure~\ref{fig:sn_distr1}.} 
\label{fig:sn_temp_redshift_equator}
\end{figure}

The GRB histogram (Figure~\ref{fig:grb_temp_redshift}) is quite random 
and shows no temperature excess. For this histogram  
 $\overline{T}_{\rm GRB}=+1.3\pm5.7~\mu{\rm K}$, which is statistically 
insignificant and which is what one 
would expect [the average number of GRBs in each redshift bin was 6.4 for 
the redshift range $z \in(0,4)$]. One could attribute the 
difference between the two histograms to the differences between 
the GRB and SN distributions across the sky:  the GRB sample is uniform,
whereas the SN sample contains fields with high source densities,
corresponding to specialised surveys of extragalactic sources.
Figure \ref{fig:sn_distr1} shows the sky distribution 
of the SNe used here for calculating 
the temperature-redshift histogram. There is a distinctive strip 
seen in this map that contains a larger number of SNe. It corresponds
to the SDSS Supernova Survey conducted for a narrow equatorial zone of 
$\sim 2.5^\circ$-width \citep{frieman08}. In order to check whether the 
excess of the WMAP pixel temperatures is due to this SN subsample, we have
calculated a separate $T_{\rm SN}/z_{\rm SN}$-histogram containing 
only the SNe with $|\delta|<1.5^\circ$
(shown in Figure~\ref{fig:sn_temp_redshift_equator}), and we have found that 
this histogram reveals a statistically-significant positive shift of the 
average temperature for low redshifts  
$\overline{T}_{\rm SN}^{(0.0,0.4)}=+11.1\pm 1.4~\mu{\rm K}$,
whereas for higher redshifts the shift is not
significant:
$\overline{T}_{\rm SN}^{(0.5,1.0)}=+3.6\pm 4.7~\mu{\rm K}$,
which indicates that the found temperature shift
for higher redshifts comes from the rest of the SN sample,
which is distributed more uniformly across the sky than the
SDSS Supernova Syrvey.
   
In order to distinguish between possible sources of 
the detected anomaly we have calculated 
similar histograms for three different WMAP frequency bands, 
separately for the Q, V and W bands (keeping in mind that,
for instance, the Sunyaev-Zeldovich effect 
is energy-dependent). These three histograms are similar 
to each other and to the histogram shown in Figure~\ref{fig:sn_temp_redshift}, 
their linear regression slopes varying only slightly:
$+27.4\pm6.9$, $+35.6\pm7.6$ 
and $+32.2\pm7.8$ for the Q, V and W bands, respectively. 
This means that a larger sample of SNe (which might become available 
in the future) is needed to get a more clear picture about 
the nature of the detected correlation. 
A similar band-independent correlation between WMAP 
temperatures and large-scale structures traced by galaxies
was found by \cite{goto12} for the Wide-field Infrared Survey 
Explorer data, which was attributed by 
the authors to the integrated Sachs-Wolfe effect.

\begin{figure}
\vspace{-0.5cm}
\hspace{-0.5cm}
\epsfig{figure=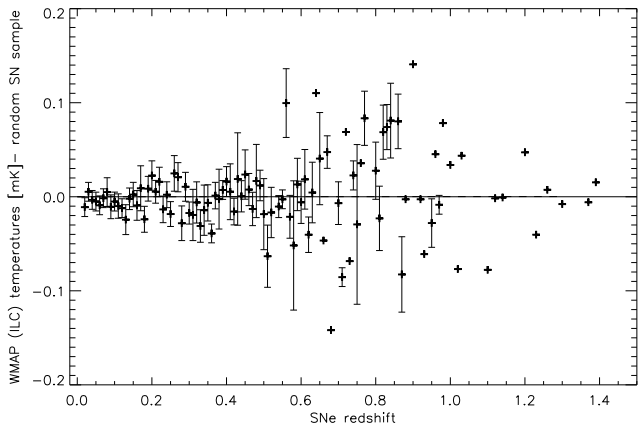,width=9.0cm}
\vspace{-0.2cm}
\caption{Histogram $T_{\rm SN}/z_{\rm SN}$ for the randomised sample
of SNe (to be compared with the histogram shown in Figure~\ref{fig:sn_distr1}.} 
\label{fig:sn_temp_redshift_rand}
\end{figure}

Yet another check ensuring that the seen correlation is not due to
a systematic error consists in randomising the SN positions on 
the sky. For this purpose, we have used the same SN sample
with assigned random coordinates, while
maintaining their redshift-distribution, as well as their higher density  
in the equatorial zone corresponding to the SDSS subsample. 
The histogram for this randomised SN locations (Figure~\ref{fig:sn_temp_redshift_rand})
does not reveal any significant temperature anomalies, for both
low ($\overline{T}_{\rm SN_{rand}}^{(0.0,0.4)}=-1.7\pm 1.8~\mu{\rm K}$)
and high ($\overline{T}_{\rm SN_{rand}}^{(0.5,1.0)}=+9.3\pm6.6~\mu{\rm K}$)
redshift ranges, with the average for the whole SN sample being
$\overline{T}_{\rm SN_{rand}}=-0.3\pm 1.6~\mu{\rm K}$. So,
the found anomaly is likely to be real.  

\section{Discussion}

In our view, this anomaly
can be interpreted in a number of different ways, of which
we would like to mention the following:

\begin{itemize}
\item[-]
it might correspond to the secondary CMB 
anisotropies caused by dust associated with clusters and 
superclusters of galaxies at moderate redshifts $z \simeq 1$;  
  
\item[-]
it could be related to secondary and tertiary ionisation 
of the foreground media during the post-reionisation epoch; 

\item[-]
it could be caused by large-scale inhomogeneities in matter distribution, 
such as voids, filaments or walls, which could alter the energies of the primordial
CMB photons through the integrated Sachs-Wolfe and Rees-Sciama effects.

\end{itemize}

These effects are broadly discussed in the literature,
so let us just briefly outline how they 
contribute to the CMB anisotropies.

{\it Dust contribution.}\,
The SN and GRB host galaxies contain dust whose thermal
radiation contributes to the CMB. 
It is known that
dust in spiral galaxies has a wide range of temperatures
\citep{rowan92,franceschini95}
which makes it difficult to separate CMB anisotropies 
caused by dust from the primordial CMB anisotropies.
Recent Hershel submillimiter observations have shown 
the presence of large dust masses in high-redshift galaxies
\citep{santini10}. 
Similarly, large dust masses have been measured
in local universe galaxies 
\citep{galliano05,bendo06,galametz09,grossi10,ohalloran10}. 
So, indeed, there is a possibility that dust from galaxies at all redshifts 
contributes to the CMB temperature fluctuations, which, in turn, would
result in a positive correlation between CMB temperatures and overdensities 
of galaxies  -- see, e.g., \citet{ho08}. 
The possible impact of dusty galaxy clusters 
on cosmological parameter estimates was also discussed by 
 \citet{serra08}, \citet{dunkley11} and \citet{millea12}.

{\it Ionised gas.}\,
Another possible source of secondary anisotropies in
the CMB power spectrum is the Thomson scattering on free electrons
in regions of ionised gas with bulk peculiar velocities  
produced at the epochs of reionisation \citep{mcquinn05,mcquinn10}
and post-reionisation. 
Local variations in the ionised gas increase 
temperature anisotropies through the kinetic 
Sunyaev-Zeldovich effect,
the primordial temperature perturbations
being damped as $e^{-\tau}$, where $\tau$ is the  Thomson scattering optical
depth.

The WMAP best-fitting model for the large-scale polarisation anisotropy
favours a mean redshift of reionisation of $10.4 \pm 1.4$ 
\citep{bennett03,komatsu09} with 
$\tau \approx 0.08$ \citep{dunkley09}. 
The observed spectra of quasars indicate a rapid change in the ionising 
background at the redshift $z\approx 6$, so that reionisation is 
completed at that epoch. 
By contrast, there are many uncertainties regarding the redshift
corresponding to the beginning of reionisation. 
The polarisation anisotropies models are based on 
a number of assumptions regarding the density distribution 
of the baryons, quasar physics, etc., much of which  are unknown.
If reionisation occurred gradually 
over a period of time, then it started as long ago as $z=30$ 
\citep{kogut03}, so that the values of $\tau = 0.08$ or even
 $\tau = 0.17$ \citep{kogut03,spergel03} is likely to be 
underestimated. As was shown by \citet{gnedin04}, 
one can construct {\it a posteriori} models that produce
a large value of the Thomson scattering optical depth
(up to $\tau=0.2$) consistent with the WMAP measurements.   
Moreover, the reionisation history could be much more  
complex than is usually thought 
\citep{barkana01,miralda03,bromm04,mcquinn07}.
According to \citet{wyithe03,cen03,haiman03,bromm04},
there might have been more than one reionisation period, 
with the first 
reionisation occurring at $z \approx 20$ due to massive population III stars, 
followed by a period of partial recombination, and, finally, the 
reionisation at $z \approx 6$.
Giant blobs of gas are known to surround galaxies and galaxy clusters
\citep{francis96,steidel00,matsuda04,hayes11}. Some of
partially ionised blobs remaining after the end of the reionisation 
epoch at $z=6$ could screen and dampen the CMB fluctuations along certain 
directions at much lower redshifts 
(this might be yet another possible explanation 
of the extreme Cold Spot anomaly).

Besides the afore-mentioned damping of the primordial fluctuations,
there exists an effect caused by inhomogeneous reionisation 
which leads to the generation of new temperature fluctuations 
on small angular scales 
\citep{babich06,babich07,holder07} due to the bulk motion of 
electrons in overdense regions \citep{ostriker86}. The patchiness
of reionisation remains theoretically uncertain and observationally
unconstrained, whereas its contribution to the CMB fluctuations 
is likely to be comparable to, or larger than, that of the Thomson scattering 
effect. 

The upper limits of $\sim 6 \, \mu{\rm K}$ on the CMB anisotropies due to the 
secondary thermal and kinetic Sunyaev-Zeldovich effects were estimated by 
\citet{leuker10} and \citet{shirokoff11}.
These are smaller than the amplitude of the anomaly we have reported here.

{\it Large-scale structure.}\,
Both CMB anisotropy and large-scale structures of the universe 
(mapped by galaxy surveys) are related to the genesis and evolution 
of primordial perturbations and to the cosmological model and its parameters. 
However, due to the existence of large-scale structures in the universe 
(high-density knots, filaments, walls etc.) there are gravitational potential 
wells all over the sky which distort the CMB power spectrum on large angular 
scales through the integrated  Sachs-Wolfe and Rees-Sciama effects. The existence of 
correlations between WMAP maps and large scale structures is well-known  
\citep{crittenden96,kinkhabwala99,boughn04,cabre06,stefanescu10},
so the correlation which we are reporting here is likely to be of a 
similar kind. \cite{granett08} measured the amplitude of the temperature 
deviations due to the integrated Sachs-Wolfe effect for galaxy superclusters 
in SDSS, which was found to be of of the order of $+9.6\pm 2.2~\mu{\rm K}$. This 
is in a good agreement with the reported here anomaly for low redshifts:
$\overline{T}_{\rm SN}^{(0.0,0.4)}=+8.6\pm 1.3~\mu{\rm K}$,
whereas the larger anomaly for higher redshifts, 
$\overline{T}_{\rm SN}^{(0.5,1.0)}=+29.9\pm4.4~\mu{\rm K}$, 
exceeds the theoretical amplitudes for the integrated Sachs-Wolfe effect.
This should be studied in more detail in the future.

As we have seen, many factors of the foreground contamination 
could distort the primordial angular power spectrum
of CMB fluctuations. 
All these effects create a rather thick patchy screen between the 
epoch of last-scattering and the current epoch. This screen  
might be responsible for a great deal of the observed CMB anisotropy.

\section{Conclusions}
The correlation between the SNe redshifts  
and WMAP temperatures indicates
that the WMAP data are still contaminated by extragalactic
foreground, despite being thoroughly cleaned from known
foreground sources.  
Thus, further studies 
are important 
in order to clarify the question concerning the fraction of the 
primordial fluctuations remaining in the CMB maps, as
the knowledge of the power spectrum of these fluctuations 
is pivotal for the calculation
of the parameters of the standard cosmological model.

\section*{acknowledgements}
\noindent
The research has made use of the following archives:
the Legacy Archive for Microwave Background Data Analysis
operated by NASA's Goddard Space Flight Center;
the Hierarchical Equal Area iso-Latitude Pixelisation (HEALPix) software
packages maintained by the Jet Propulsion Laboratory,
California Institute of Technology, Pasadena;
the list of GRB sources maintained by J.Greiner, Max-Planck
-Institut f\"ur extraterrestrische Physik, Garching;
the Leicester Database and Archive Service (LEDAS) operated by
the University of Leicester;
the GRB pseudo-redshift catalogue maintained by 
A. P\'elangeon and J.-L. Atteia at Laboratoire
d'Astrophysique de Toulouse-Tarbes, Universit\'e 
de Toulouse; and the Supernovae Catalogue maintained by 
Sternberg Astronomical Institute, Moscow State University.
The authors appreciate the helpful comments
made by Dr. L.V. Morrison.  
One of the authors (V.V. Orlov) is thankful 
to the Russian President Grants Council for  
the State Support of the Leading 
Scientific Schools (the Grant NSCH-3290.2010.2).

\end{document}